\documentclass[twocolumn,twoside,slac_two]{revtex4}
\usepackage{graphicx}
\usepackage{fancyhdr}
\begin{document}

\title{Population Synthesis of Radio Pulsars in the Fermi Era}

\author{P. L. Gonthier}
\affiliation{Hope College, Holland, MI 49423, USA}

\author{E. Nagelkirk}
\affiliation{Wayne State University, Detroit, MI 48202, USA}

\author{M. Stam}
\affiliation{Georgia Institute of Technology, Atlanta, GA 30332, USA}

\author{A.K. Harding}
\affiliation{NASA Goddard Space Flight Center, Greenbelt, MD 20771, USA}

\begin{abstract}
We present results of our pulsar population synthesis of normal pulsars
from the Galactic disk using our previously developed Monte-Carlo code. 
From our studies of observed radio pulsars that have clearly
identifiable core and cone components, in which we fit the polarization
sweep as well as the pulse profiles in order to constrain the viewing
geometry, we develop a model describing the luminosity and ratio of
radio core-to-cone peak fluxes.  In this model, short period pulsars are
more cone-dominated.  We explore models of neutron star evolution with
and without magnetic field decay, and with different initial period
distributions.  We present preliminary results including simulated
population statistics that are compared with the observed radio pulsar
population.   The evolved neutron star populations resulting from this
simulation can be used to model distributions of $\gamma$-ray pulsars for
comparison to {\it Fermi} results.

\end{abstract}

\maketitle

\thispagestyle{fancy}

\section{INTRODUCTION}
NASA's new flagship in $\gamma$-ray astronomy, the Fermi Gamma-Ray Space
Telescope ({\it Fermi}) launched in June 2008, has already dramatically
improved our understanding of $\gamma$-ray emission from pulsars.  In
its first year of operations, {\it Fermi} has discovered over 40 new pulsars
above 20 MeV in photon energy, superseding the $\gamma$-ray pulsar
database of six provided by its predecessor, the EGRET instrument aboard
the Compton Gamma-Ray Observatory.  Over a dozen of these have been
discovered through blind searches in $\gamma$ rays, most of which
display, as yet, no evidence of a radio signal.  In addition, {\it Fermi} has
identified nearly a dozen millisecond pulsars that have been observed in
the radio. 

With the wealth of new data from {\it Fermi} at hand, we have the
first real opportunity in decades to finally understand the high-energy
emission and acceleration in pulsar magnetospheres.  Making full use of
this opportunity will require more detailed modeling of different
emission mechanisms and geometries, and comparing model predictions with
the well-defined trends in the observations should be very productive. 
In addition to better defining pulsar $\gamma$-ray emission, it will be
possible also to constrain the radio emission location and geometry. 
The large collection of radio-quiet/radio-weak $\gamma$-ray pulsars are
defining the viewing angles where we are just crossing the outer edge of
the radio beam or missing it altogether.  Modeling both the radio beam
and $\gamma$-ray beam geometry together (as presented in companion paper
Pierbattista et al. and was recently done for millisecond pulsars by
Venter, Harding \& Guillemot \cite{Venter}.) will be able to provide
meaningful constraints of beam pattern and emission altitude range.

\section{MAGNETIC FIELD AND PERIOD BIRTH DISTRIBUTIONS}

\subsection{Case A Ð Magnetic field decay}
We simulate the population of normal radio pulsars using two different
sets of assumptions.  Case A assumes that the pulsar spin-down can be
mimicked through the incorporation of the decay of the magnetic field
with a decay constant of 2.8 Myr following the study of Gonthier et al.
\cite{Gont1}.  While we do not advocate that this result provides clear evidence
for field decay, we find that this method allows one to incorporate an
alternative to the standard vacuum, dipole spin-down, as for example in
the work of Contopoulos \& Spitkovsky \cite{Cont}.  The initial magnetic field is
assumed to be described by two log-normal distributions given by the
expression
\begin{equation}\label{eq:pb}
P\left( {\log B_{\rm{o}} } \right)= \sum\limits_{i = 1}^2 {A_i e^{ - \left( {\log B_{\rm{o}}  - \log B_i } \right)^2  /\sigma_i^2}}
\end{equation}
with the following parameters indicated in Table~\ref{tab-1}.
\begin{table}[b]
\begin{center}
\caption{Magnetic field distribution parameters}
\begin{tabular}{|c|c|c|c|}
\hline $\mathbf i$ & $\mathbf A_i$ & $\mathbf {\log (B_i)}$ & $\mathbf{ \sigma_i}$
\\
\hline 1 & 0.6 & 12.5 & 0.65 \\
\hline 2 & 0.3 & 13.0 & 0.80 \\
\hline
\end{tabular}
\label{tab-1}
\end{center}
\end{table}
The initial period distribution for this case A is given by a Gaussian with a mean $\hat P_o$ and width $\sigma_{P_o}$ in
\begin{equation}\label{eq:pos}
\begin{array}{c}
 P\left( {P_o } \right) \propto e^{ - \left( {P_o  - \hat P_o } \right)^2 /\sigma _{P_o }^2 }  \\ 
 \hat P_o  = 300{\rm{ ms}} \\ 
 \sigma _{P_o }^{}  = 300{\rm{ ms}} \\ 
 \end{array}
\end{equation}

\subsection{Case B Ð No field decay}
Since the short decay constant of 2.8 Myr assumed for case A is physically unrealistic for magnetic field decay, we develop a no-field decay model exploring a radio luminosity law that is proportional to the square root of the spin-down power as suggested by the study of Faucher-Gigure \& Kaspi \cite{Fauc}.  However in their study, they did not include the geometry of the radio beam but rather used a standard beaming model to account for the average beam characteristics.  With our set of assumptions defining the radio beam geometry and luminosity, we are unable to reproduce the observed $\dot P - P$ distribution.  In order to find reasonable agreement, we correlate the initial period distribution with the initial magnetic field, which remains constant during the pulsar spin-down.  The distribution is described with a single log-normal $B$ distribution and a correlated Gaussian $P_o$ distribution by the expression 
\begin{widetext}
\begin{equation}\label{eq:bpo}
\begin{array}{c}
{\scriptstyle  P\left( {B,P_o } \right) \propto }
{\scriptstyle  \exp \left\{ { - \left[ { \left( {\log B - \mu _{\log B} } \right)^2 /\sigma _{\log B}^2  + \left( {P_o  - \mu _{P_o } } \right)^2 /\sigma _{P_o }^2 
   - 2\rho \left( {P_o  - \mu _{P_o } } \right)\left( {P_o  - \mu _{P_o } } \right)/\left( {\sigma _{\log B}^{} \sigma _{P_o }^{} } \right) \hfill} \right]  /\left[ {2\left( {1 - \rho ^2 } \right)} \right]} \right\} }
 \end{array}
\end{equation}
\end{widetext}
with the following parameters
\begin{equation}\label{eq:bpos}
\begin{array}{c}
 \mu _{\log B}  = 12.9 \\ 
 \sigma _{\log B}^{}  = 0.7 \\ 
 \mu _{P_o }  = 200\ {\rm{ ms}} \\ 
 \sigma _{P_o }^{}  = 100\ {\rm{ ms}} \\ 
 \rho  =  - 0.6 \\ 
 \end{array}
\end{equation}

\begin{figure*}[t]
\centering
\includegraphics[width=170mm]{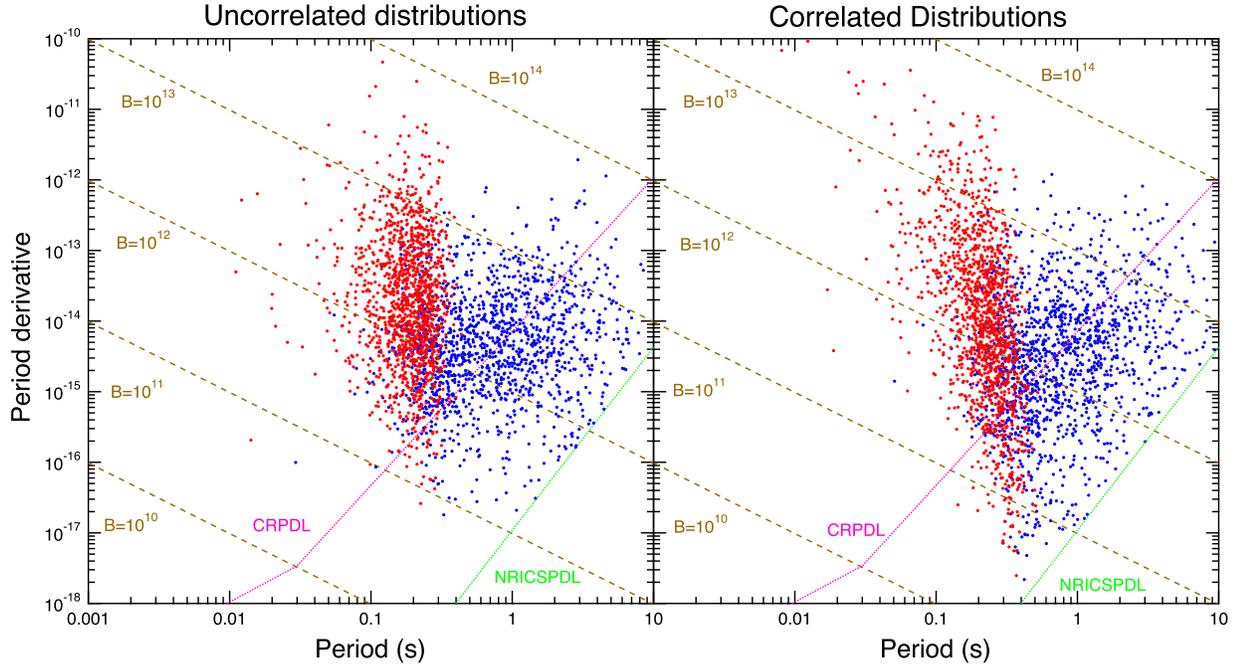}
\caption{ $\dot P - P$ distribution of pulsars at birth (red) and at present (blue) illustrating the effect of correlating the magnetic field with the initial period using the same distributions in both cases.  On the left the magnetic field distribution is the above distribution without the cross term while on the right they are correlated with the coefficient $\rho=-0.6$.} \label{f-1}
\end{figure*}

In Figure~\ref{f-1}, we display the effect of this correlated distribution on the present-day distribution of pulsars in the  $\dot P - P$ diagram.  The distribution on the left shows the initial distribution (red) and the present-day distribution (blue) of simulated pulsars assuming there is no correlation between the initial period and the magnetic field, while the distribution on the right shows the corresponding distributions with the correlation parameter of $\rho=-0.6$ as indicated above (Eq.~\ref{eq:bpos}).  The distribution of detected pulsars resembles an upside-down pear shape.  We find that it is not trivial to reproduce the narrowing of the shape of the distribution at the smallest  with a paucity of pulsars to the left side (shorter period) of this narrowing around a period of about 0.7 seconds.  Correlating the initial period with the magnetic field does seem to provide the necessary features to account for the observed distribution.  We do realize that this correlation scheme is difficult to probe observationally. In both cases A and B, the minimum initial period in the simulation is set to 1.3 ms.

\section{RADIO BEAM GEOMETRY AND LUMINOSITY}
\subsection{Core Ð Cone Model}
As in our previous study \cite{Gont1}, we assume a beam model that is characterized by a central core beam and a single cone beam within the open-field volume with the cone beam decreasing in intensity near the last open field surface.  The core beam is assumed to be a traditional Gaussian centered along the magnetic axis with a 1/e characteristic width \cite{Arzou}. 
\begin{equation}\label{eq:core}
\rho _{{\rm{core}}}  = 1^ \circ  .5P^{ - 1/2} 
\end{equation}
Following the work of Kijak \& Gil \cite{Kija1,Kija2} we describe the hollow conal beam with a Gaussian having a characteristic beam radius given by
\begin{equation}\label{eq:cnre}
\begin{array}{c}
 \rho _{{\rm{cone}}}  = 1^ \circ  .24r_{KG}^{1/2} P^{ - 1/2}  \\ 
 r_{KG}  = 40\nu _{{\rm{GHz}}}^{{\rm{ - 0}}{\rm{.26}}} \dot P_{ - 15}^{0.07} P^{0.30}  \\ 
 \end{array}
\end{equation}
where $r_{KG}$ is the emission altitude in stellar radii.  The characteristic beam radius corresponds to a conal beam opening angle at the 0.1\% of the peak intensity of the observed profile.

\subsection{Radio Luminosity}
Following the standard candle assumption of radio pulsars, we use a luminosity model with the same form as used in the work of Arzoumanian, Chernoff \& Cordes \cite{Arzou} given by the expression
\begin{equation}\label{eq:lumo}
L_{{\rm{radio}}}  = \frac{{66250}}{{R_f }}P^\alpha  \dot P_{15}^\beta  {\rm{ mJy}} \cdot {\rm{ kpc}}^2  \cdot {\rm{ MHz}}
\end{equation}
where $R_f$  is a reduction factor that adjusts the birth rate for a given pair of $\alpha$ period and $\beta$ period derivative exponents.   $\dot P_{15}$ is the period derivative in units of  $10^{-15}\ \rm{s\cdot s^{-1}}$.  The exponents aid in reproducing the observed  $\dot P - P$ distribution, and the reduction factor is adjusted to provide a neutron star birth rate of about 2.1 per century \cite{Tamm}.  Our two cases A and B described above require different sets of coefficients shown below in Table~\ref{tab-2}.

\begin{table}[t]
\begin{center}
\caption{Radio luminosity parameters}
\begin{tabular}{|c|c|c|c|}
\hline \textbf{Case} & $\bf R_f$ & $\bf {\alpha}$ & $\bf{ \beta}$
\\
\hline A & 1.7 & -1.0 & 0.35 \\
\hline B & 2.0 & -1.5 & 0.50 \\
\hline
\end{tabular}
\label{tab-2}
\end{center}
\end{table}
Following the suggestion \cite{Fauc}, we constrain the radio luminosity for the no-field decay case B to be proportional to the square root of the spin-down power to explore if the need for field decay stems from our choice of the functional form of the radio luminosity in our field-decay case A.  Despite implementing the radio luminosity for case B, following the suggestion of Faucher-Gigure \& Kaspi \cite{Fauc}, we find that to reproduce the  $\dot P - P$ distribution we need to correlate the initial period with the magnetic field distribution as discussed in Section 2 for case B.  The correlation scheme appears to have a somewhat similar effect as field decay, both of which are most likely mimicking a pulsar spin-down that deviates from that of the traditional vacuum dipole.  New theoretical treatments are encouraging (see the contribution of Spitkovsky these proceedings) and may provide better insight on the pulsar spin-down.

\begin{figure*}[t]
\centering
\includegraphics[width=170mm]{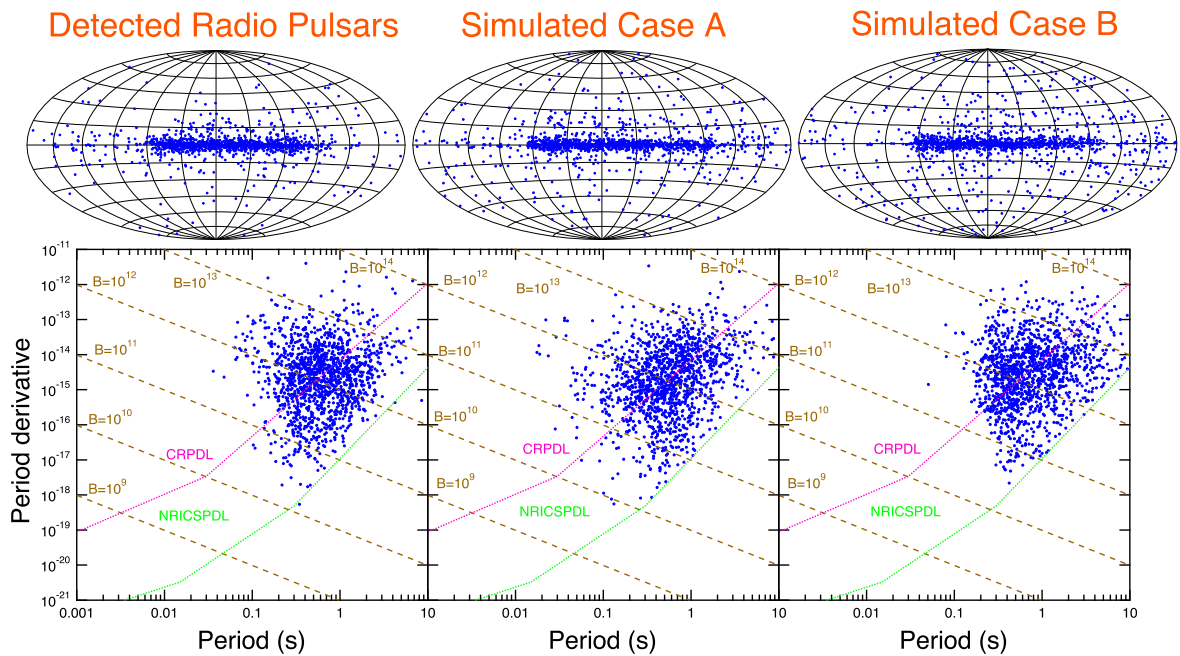}
\caption{ Aitoff plots (above) and  $\dot P - P$ diagrams (below) for detected (left), simulated case A (center) and case B (right).  The brown dashed lines represent lines of constant magnetic field (indicated) assuming dipole spin-down.  The pink and green dotted curves are the curvature radiation and non-resonance inverse Compton scattering pair death lines, respectively.} \label{f-2}
\end{figure*}

\subsection{Core-to-cone peak flux model}
A core-to-cone peak flux ratio as used in the work of Arzoumanian, Chernoff \& Cordes \cite{Arzou} is required to partition the luminosity between the core and cone components of the radio beam.  We studied (Gonthier et al. \cite{Gont2}) a group of 20 pulsars in the EPN database primarily from the observations of Gould \& Lyne \cite{Goul} whose pulse profiles manifested three peaks.  Using the Rotating Vector Model, we fit the polarization sweep to obtain the maximum rate of change of the position angle at the inflection point $(d\psi/d\phi)_{\rm max}$.  This parameter is equal to the ratio of the sine of the magnetic inclination angle and the sine of the impact angle.  Using this parameter for a particular pulsar, we fit the profile shape allowing for the determination of the viewing geometry and thus obtain the ratio of the core-to-cone peak fluxes for those 20 pulsars within the assumed beam geometry and luminosity model.  We find that we can describe this ratio with the following broken power law.
\begin{equation}\label{eq:peak}
r_{{\rm{peak}}}  = \left\{ \begin{array}{l}
 25P^{1.3} \nu ^{ - 0.9} ,{\rm{\  for\ }}P < 0.7{\rm{s}} \\ 
 4P^{ - 1.8} \nu ^{ - 0.9} ,{\rm{\  for\ }}P > 0.7{\rm{s}} \\ 
 \end{array} \right.
\end{equation}
where $\nu$ is the observing frequency in MHz.  In this model, short period pulsars are much less core dominated than in the model used in the study of Arzoumanian, Chernoff \& Cordes \cite{Arzou}.  In fact, when the period is less than 50 ms, the profile is cone dominated.  This model described in Eqn.~\ref{eq:peak} appears somewhat consistent with several recent findings. Crawford et al. \cite{Craw1,Craw2} studied a number of young pulsars, finding that the profiles have large linear polarization and little circular polarization, suggesting that the emission is cone dominated.  More recently Johnston \& Weisberg \cite{John} have a similar conclusion from polarization studies of 14 young pulsars, and Weltevrede \& Johnston \cite{Welt} found that the fractional polarization of pulsars is high $\dot E$ for pulsars with large spin-down powers  and low $\dot E$ for low  with the transition taking place near the curvature radiation pair death line.

\begin{figure*}
\centering
\includegraphics[width=170mm]{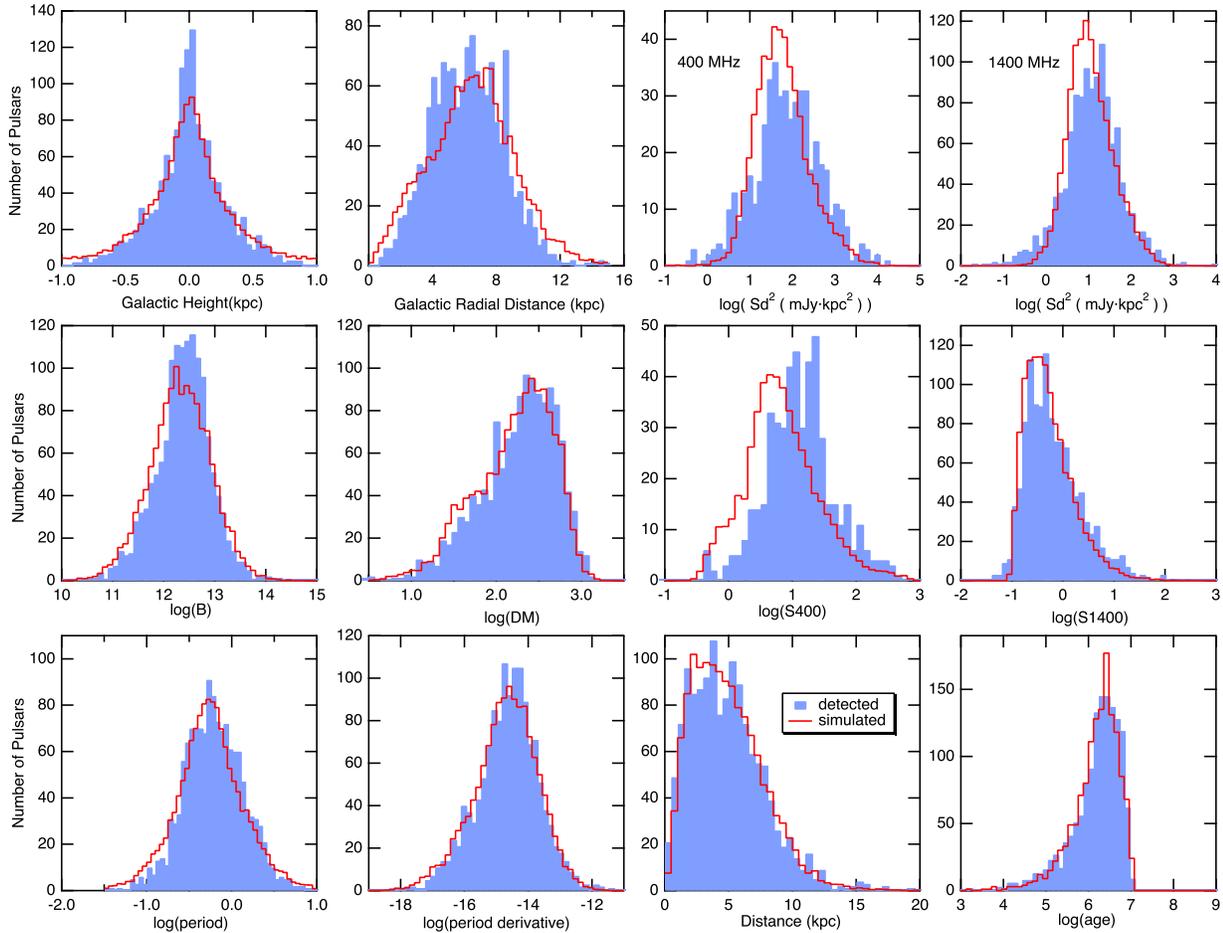}
\caption{ Histograms of the indicated pulsar characteristics.  The solid blue histograms represent the detected radio pulsars while the open red histograms represent the simulated pulsars for case A assuming magnetic field decay with a time constant of 2.8 Myr.} \label{f-3}
\end{figure*}

\section{RESULTS}
In Figure~\ref{f-2}, we present results of our population simulation for the two cases A and B, and the comparison with 1380 detected pulsars by a group of ten radio surveys including the Parkes Multibeam and Swinburne Intermediate Latitude surveys.  Above in Figure~\ref{f-2} are the Aitoff plots of the detected pulsars (left), case A (field decay) (middle) and case B (no field decay) (right) and corresponding  $\dot P - P$ diagrams below. Comparisons of the Aitoff plots suggest that the simulation is producing too many pulsars at high latitudes outside of the Galactic plane especially for case B.  The distribution of detected pulsars in the  $\dot P - P$ diagram is broad at high fields or large  $\dot P$'s and narrows considerably at low fields or small $\dot P$'s .  This trend in the detected distribution is more or less reproduced in the simulated distributions for the field decay case A and the no-field decay case B.  In order to reproduce the narrowing of the distribution at low fields for case B, we found the need to correlate the initial period with the magnetic field as discussed in Section 2.  Without this correlation, the distribution at low fields is too broad with many more pulsars with small   $\dot P$'s and short periods.  The comparisons are not optimal, as the parameter space was not fully searched, which is one of the drawbacks of a Monte Carlo method.  There are several interrelated parameters that affect the distribution and further simulation runs might lead to improved agreement. 

\begin{figure*}
\centering
\includegraphics[width=170mm]{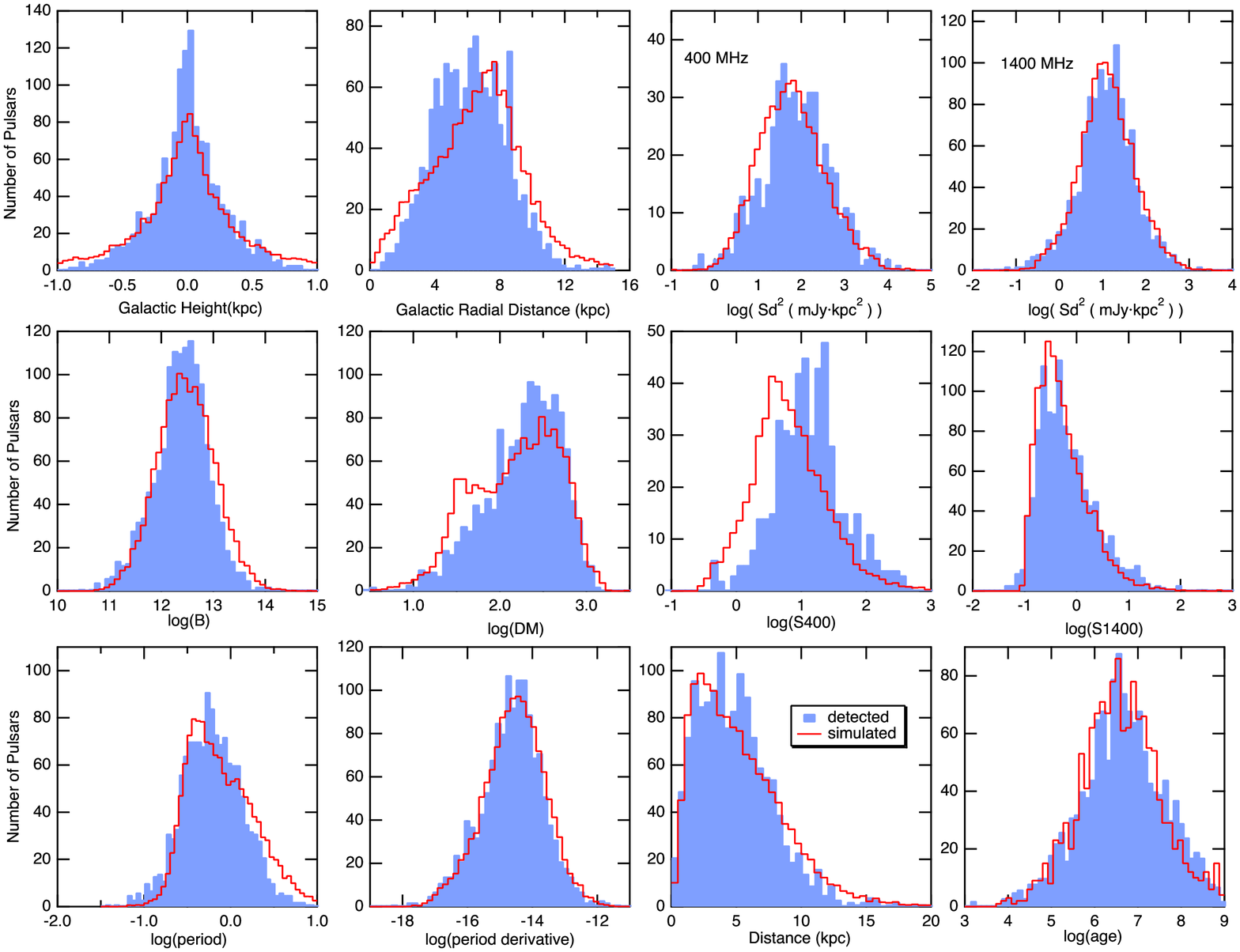}
\caption{ Histograms of the indicated pulsar characteristics.  The solid blue histograms represent the detected radio pulsars while the open red histograms represent the simulated pulsars for case B assuming constant magnetic field but correlating the initial magnetic field distribution with the initial period distribution as given by Eqn.~\ref{eq:bpo}.} \label{f-4}
\end{figure*}

Further comparisons of single pulsar characteristics are made in Figures 3 and 4 below by displaying histograms of various parameters (indicated) associated with the detected and simulated pulsars.  Directly observed parameters are the period, period derivative, dispersion measure (DM), fluxes at 400 MHz (S400) and 1400 MHz (S1400).  The distance of the detected pulsar is the best-estimated distance from the ATNF catalog, which if obtained from the DM, the NE2001 electron density model of Cordes \& Lazio \cite{Cord} is used to obtain the pulsar distance.  We use this model to obtain the pulsar DM from the distance and Galactic location of the simulated pulsar.  Having the distance, we obtain the pseudo-luminosity $Sd^2$ of the pulsar by multiplying the flux by the distance squared.  The Galactic height and radial distance are obtained using the distance and their Galactic location assuming that the Sun is located at 8.5 kpc from the Galactic center.  For case A, assuming field decay, the age of the detected and simulated pulsars are recalculated assuming the decay constant of 2.8 Myr used in the simulation and zero initial period.  Though not quite the actual age, both simulated and detect pulsars are treated in the same manner.  As a result, the age distribution for this case A reflects a shorter mean age than the one calculated from the characteristic age   $P/2 \dot P$.  For the pulsars in case B that assumes a constant magnetic field, the ages of the pulsars are the characteristic ages for both detected and simulated pulsars.  While the simulation tracks the actual age of the pulsar for both case A and B, we cannot compare it to the ages of detected pulsars without knowing their initial periods. 

The Galactic scale height distributions of the simulated pulsars for both cases are broader than the one for the detected pulsars, which was suggested previously in the Aitoff projections in Figure~\ref{f-2}.  The scale height in the initial birth distribution used in the simulation of 75 pc is the same one that was used in previous study \cite{Gont1} of normal pulsars and was not varied in the present simulations.  The distributions of Galactic radial distance and the distance to Earth are well reproduced in case A but not quite as well in case B with more pulsars closer to us than those detected, skewing the Galactic radial distance distribution to larger distances near 8 kpc.  The bump appearing in the simulated dispersion measure distribution at low DM for case B is a reflection of the distance distribution having too many nearby pulsars.  The distance distribution is especially sensitive to the radio luminosity. The larger than observed number of nearby pulsars suggest that the luminosity in the simulation is too small.  However, the pulsar birth rate provides an additional constraint.  The simulated birth rates were 2.1 and 0.9 per century for cases A and B, respectively.  Increasing the luminosity further for case B would have improved the distance distribution, but would have lowered the birth rate to an even smaller value.

The period and period derivative distributions are well reproduced for both cases A and B with a few too many short period pulsars in case A and a few too many long period pulsars in case B.  Therefore the derived magnetic field and age distributions are well reproduced.  The overall distributions of the directly measured and derived characteristics of the pulsars are better reproduce by case A than by case B.

\section{CONCLUSION}
From our simulations, using two sets of general assumptions, we find that the standard vacuum dipole spin-down of a pulsar is not able to reproduce the distribution of detected pulsars in the  diagram, and either magnetic field decay (case A) or correlating the initial period with the magnetic field with no decay (case B) is necessary to qualitatively reproduce the upside-down pear-shape distribution of the detected pulsars.  This result suggests that an alternative spin-down model maybe required.  The evolved groups of neutron stars for both of these cases provide populations of $\gamma$-ray pulsars that are compared to the characteristics of {\it Fermi} pulsars (see the contribution by Pierbattista, Grenier, Harding \& Gonthier, these proceedings).

\begin{acknowledgments}
The authors wish to express their gratitude for the generous support of the Michigan Space Grant Consortium, of the National Science Foundation (REU \& RUI) and the NASA Astrophysics Theory and Fundamental Program.
\end{acknowledgments}


\end{document}